\documentstyle[12pt]{article}
\begin{document}
\pagestyle{myheadings}
\markright{Classical action for a Bianchi $VI_h$ model
\hspace{0.75 in} gr-qc/9903073}
\date{18 March 1999}
\title{The classical action for a Bianchi $VI_{h}$ model
}
\author{R. Michael Jones
\thanks{e-mail: R.Michael.Jones@Colorado.edu}
 \\CIRES, University of Colorado\\ 
Boulder, Colorado 
80309-0216
, U.S.A.
}
\maketitle
\begin{abstract}
	An estimate for the classical action $I_{classical}$ for a
Bianchi $VI_h$ homogeneous spatially closed model with $h=-1/9$
is given by
\begin{eqnarray*}
\frac{I_{classical}}{\hbar} & \approx & \frac{3}{5}i\pi^2
(1+\alpha_0) \left[\left(\frac{r(t)}{L^*}\right)^2
\left(\frac{r(t)}{r_{m2}}\right)^{1/2} 
\right. \\
& - & 2
\left.
\left(\frac {r(t)} {r_1} \right) 
\left(\frac {r(t)} {r_{m2}} \right)^{3/2} 
\left(\frac {L^*} {r_0} \right)^2 
\left(\frac{b}{8a_0^3 L^{*2}}\right)^2 
\right] ,
\end{eqnarray*}
where $b$ and $a_0$ are parameters of the model, $b$ is zero if
and only if the relative rotation of inertial frames and matter
is zero, $\alpha_0$ is an initial value of an anisotropy of the
expansion rate at $r=r_0$, $L^*$ is the Planck length, and
$r_{m2}$ is a constant of integration that gives the maximum size
of the universe for the isotropic ($b=0$, $\alpha_0=0$) case.  

	It is assumed that the equation of state is
$p=(\gamma-1)\rho$, where $p$ is pressure and $\rho$ is density. 
It is assumed that $\gamma$ has a constant value of $\gamma=4/3$
(to represent a relativistic early universe) for $r<r_1$ and a
constant value of $\gamma=1$ (to represent a matter-dominated
late universe) for $r>r_1$.  The approximation is valid for $b$
small enough that $\left| I_{classical} [b] - I_{classical} [b=0]
\right| < \hbar$ .

	An explanation for why our inertial frame seems not to
rotate relative to the stars is found in a straightforward
application of semiclassical approximations to quantum cosmology. 
Application of a saddlepoint approximation leads to the result
that only those classical geometries whose action $I_{classical}$
satisfies $\left| I_{classical} - I_{saddlepoint} \right| <
\hbar$ contribute significantly to the integration to give the
present value of the wave function.  Using estimates for our
universe implies that only those classical geometries for which
the present relative rotation rate of inertial frames and matter
are less than about $10^{-130}$ radians per year contribute
significantly to the integration.  This is well below the limit
set by experiment.  The result depends on the Hubble distance
being much larger than the Planck length, but does not depend on
the details of the theory of quantum gravity.
\end{abstract}

\newcommand{\JournalTitle}{\em}
\newcommand{\Volume}{\bf}
\newcommand{\BookTitle}{\em}

\section{Introduction}
	Although Newton recognized that inertial frames seem not to
rotate relative to the stars, he seems to have taken that as
evidence for the existence of absolute space.  Ernst Mach
\cite{Mach72,Mach33} was probably the first physicist to
recognize that such an apparent coincidence requires an
explanation.  He based his arguments on the observation that only
relative positions of bodies are observable.  Mach further
suggested that matter might cause inertia.  

	Einstein \cite{Einstein18} tried to include what he called
``Mach's Principle'' in General Relativity, and although it is
generally agreed that matter is a source of inertia in his
theory, it is not the only source, because there are solutions
(such as empty Minkowski space) that have inertia without matter. 
Further, solutions of Einstein's field equations include
cosmologies where inertial frames rotate relative to the bulk of
matter in the universe, so that an explanation for why our
inertial frame does not rotate relative to the stars is still
needed.

	It has been suggested (e.g. \cite{Wheeler64}) that Mach's
principle be used as a boundary condition to eliminate those
solutions of the field equations that have inertia from sources
other than from matter.  Similar suggestions (e.g.
\cite{Altshuler67,Lynden-Bell67,SWG69,Gilman70,Raine75}) that
such selection might take place automatically if a satisfactory
integral formulation of Einstein's field equations might be found
would eliminate the need for explicit boundary conditions.  Other
suggestions in which gravitation (including inertia) is
represented by a theory analogous to Newtonian gravitation or
Maxwell theory (e.g. \cite{Sciama53,Lynden-Bell92}) are also
intriguing, and have the similar property that the gravitational
field would be determined solely by the matter distribution.  

	Even if we could successfully find a theory along the lines
of those mentioned above, in which matter somehow determined the
gravitational field, it might not be satisfactory.  Rather than
explain why our inertial frame seems not to rotate relative to
the stars, such a theory would simply impose that condition, and
would take away all degrees of freedom from the gravitational
field.  We are not so restrictive with the electromagnetic field,
for example.  We allow arbitrary initial values on the
electromagnetic field that are consistent with Maxwell's
equations.  We don't require that the electromagnetic field be
completely determined by charges and currents.  Just as the
electromagnetic field should be as fundamental as charges and
currents, the gravitational field should be as fundamental as
matter \cite{Earman93,Kuchar93,Raine93}.

	Allowing arbitrary initial conditions for the gravitational
field (consistent with the field equations) is inconsistent with
trying to explain why our inertial frame seems not to rotate
relative to the stars, at least on the classical level.

	On the quantum level, however, we might imagine that somehow
the selection takes place automatically through wave
interference, and this turns out to be the case.  To show that
requires calculating the action for a classical cosmology as a
function of the appropriate parameters of the model, which is the
goal here.

\section{An example from ordinary wave mechanics}
	To help explain the ideas that follow, we first consider
elementary wave mechanics as an example.  If we have an initial
single-particle state specified by an initial wave function
$<x_{1},t_{1}|\psi>$ at time $t_{1}$ then the wave function
$<x_{2},t_{2}|\psi>$ at time $t_{2}$ is \cite[p.
57]{FeynmanHibbs65}
\begin{equation}
<x_2,t_2|\psi> = \int_{-\infty}^{\infty} <x_2,t_2|x_1,t_1>
<x_1,t_1|\psi> dx_1,  \label{2.1}
\end{equation}
where $<x_{2},t_{2}|x_{1},t_{1}>$ is the propagator for the
particle to go from $(x_{1},t_{1})$ to $(x_{2},t_{2})$.  We
consider the case where the semiclassical approximation for the
propagator is valid.  That is, \cite[p. 60]{FeynmanHibbs65}
\begin{equation}
<x_2,t_2|x_1,t_1>\approx f(t_1,t_2)
e^{\frac{i}{\hbar}I_{cl}[x_2,t_2;x_1,t_1]} , \label{2.2}
\end{equation}
where $I_{cl}[x_{2},t_{2};x_{1},t_{1}]$ is the action calculated
along the classical path from $(x_{1},t_{1})$ to $(x_{2},t_{2})$. 
Thus, (\ref{2.1}) becomes
\begin{equation}
<x_2,t_2|\psi> \approx f(t_1,t_2) \int_{-\infty}^{\infty}
e^{\frac{i}{\hbar} I_{cl}[x_2,t_2;x_1,t_1] } <x_1,t_1|\psi> dx_1
. \label{2.3}
\end{equation}
Notice that because of the initial wave function we have an
infinite number of classical paths contributing to each value of
the final wave function.  

	There are two cases to consider.  In the first, $I_{cl}$ is
not a sharply peaked function of $x_{1}$.  In that case, there
will be contributions to the wave function at $t_{2}$ from
classical paths that differ greatly from each other.  

	In the second case, which we now consider, $I_{cl}$ is
sharply peaked about some value of $x_{1}$, say $x_{sp}$.  That
is, we have 
\begin{equation}
\frac{\partial}{\partial x_1}
I_{cl}[x_2,t_2;x_1,t_1]|_{x_1=x_{sp}} = 0. \label{2.4}
\end{equation}
Thus, $x_{sp}$ is a saddlepoint of the integral (\ref{2.3}), and
significant contributions to the integral are limited to values
of $x_{1}$ such that
\begin{equation}
|x_1-x_{sp}|^2 < \left|\frac{2\hbar}{\frac{\partial ^2}{\partial
x_1^2} I_{cl}[x_2,t_2;x_1,t_1]|_{x_1=x_{sp}}}\right| .
\label{2.5}
\end{equation}
If $<x_{1},t_{1}|\psi>$ is nearly constant over that range, then
we can take it outside of the integral.  A saddlepoint evaluation
of the integral then gives
{ \samepage
\begin{eqnarray}
<x_2,t_2|\psi> & \approx & f(t_1,t_2) <x_{sp},t_1|\psi> \nonumber
\\
 & & \left[\frac{2\pi i\hbar}{\frac{\partial ^2}{\partial x_1^2}
I_{cl}[x_2,t_2;x_1,t_1]|_{x_1=x_{sp}}}\right]^{1/2}
e^{\frac{i}{\hbar}I_{cl}[x_2,t_2;x_{sp},t_1]} . \label{2.6}
\end{eqnarray}
}

	We notice from (\ref{2.4}) that the momentum at $t_{1}$ at
the saddlepoint is zero.  That is,
\begin{equation}
p_1|_{x_1=x_{sp}} = 0 . \label{2.7}
\end{equation}
However, for the paths that contribute significantly to the
integral in (\ref{2.3}), there is a range of momenta, namely
\begin{equation}
|p_1^2| < 2\hbar \left|\frac{\partial ^2}{\partial x_1^2}
I_{cl}[x_2,t_2;x_1,t_1]|_{x_1=x_{sp}}\right| , \label{2.8}
\end{equation}
consistent with (\ref{2.5}) and the uncertainty relation.  

	As a check, using a special case, we consider the
free-particle propagator \cite[p. 42]{FeynmanHibbs65}
\begin{equation}
<x_2,t_2|x_1,t_1> = \left[\frac{m} {2\pi i \hbar
(t_2-t_1)}\right]^{1/2} exp \left[\frac{im(x_2-x_1)^2} {2\hbar
(t_2-t_1)}\right] , \label{2.9}
\end{equation}
and we choose 
\begin{equation}
<x_1,t_1|\psi> = <A,t_1|\psi> exp[-B(x_1-A)^2] \label{2.10}
\end{equation}
to represent a broad initial wave function.  For this case, the
integral in (\ref{2.1}) or (\ref{2.3}) can be evaluated exactly
to give
\begin{equation}
<x_2,t_2|\psi> = <A,t_1|\psi> \left[1+\frac{2B\hbar (t_2-t_1)}
{im}\right]^{-1/2} exp\left[\frac{-B(x_2-A)^2}
{1-\frac{2B\hbar(t_2-t_1)}{im}}\right] . \label{2.11}
\end{equation}
The condition that the initial wave function is slowly varying is
now
\begin{equation}
|B| \ll \left|\frac{m}{2\hbar (t_2-t_1)}\right| , \label{2.12}
\end{equation}
so that (\ref{2.11}) is approximately
\begin{equation}
<x_2,t_2|\psi> = <x_2,t_1|\psi>  \label{2.13}
\end{equation}
in agreement with (\ref{2.6}), since
\begin{equation}
I[x_2,t_2;x_1,t_1] = I_{cl}[x_2,t_2;x_1,t_1] = \frac{m}{2}
\frac{(x_2-x_1)^2}{t_2-t_1} , \label{2.14}
\end{equation}
and
\begin{equation}
x_{sp} = x_2 . \label{2.15}
\end{equation}
Notice that it is the sharply peaked action that determines which
classical paths in (\ref{2.3}) dominate the integral in this
case, not the maximum of the initial wave function.  

	The calculations can clearly be generalized to two or three
dimensions.  The main point is that whenever the initial wave
function is broad and the action of the classical propagator is
not a sharply peaked function of $x_1$, some very different
classical paths may contribute to the wave function in the final
state, even when a semiclassical approximation is valid for the
propagator.  

	When the classical action is sharply peaked as a function of
the coordinates of the initial state, however, only a narrow
range of classical paths contribute significantly to the wave
function in the final state.  This is thus a mechanism for
selecting classical paths in wave mechanics.  As we shall argue
in the next sections, this principle has broader application.
\section{Quantum cosmology}
	In the case of quantum cosmology, we have a formula
analogous to (\ref{2.1}) to give the wave function over
3-geometries $g_{2}$ and matter fields $\phi_{2}$ on a
3-dimensional hypersurface $S_{2}$.
\begin{equation}
<g_2,\phi_2,S_2|\psi> = \int <g_2,\phi_2,S_2|g_1,\phi_1,S_1>
<g_1,\phi_1,S_1|\psi> D(g_1) D(\phi_1) , \label{3.1}
\end{equation}
where $<g_{1},\phi_{1},S_{1}|\psi>$ is the wave function over
3-geometries $g_{1}$ and matter fields $\phi_{1}$ on a
3-dimensional hypersurface $S_{1}$, and
$<g_{2},\phi_{2},S_{2}|g_{1},\phi_{1},S_{1}>$ is the amplitude to
go from a state with 3-geometry $g_{1}$ and matter fields
$\phi_{1}$ on a surface $S_{1}$ to a state with 3-geometry
$g_{2}$ and matter fields $\phi_{2}$ on a surface $S_{2}$
\cite{Hawking79}.  $D(g_{1})$ and $D(\phi_{1})$ are the measures
on the 3-geometry and matter fields.  The integration is over all
initial 3-geometries $g_{1}$ and matter fields $\phi_{1}$ for
which the integral is defined.

\section{Semiclassical approximation}
	As in section 2, we want to consider the case where the
semiclassical approximation for the propagator is valid.  That
is, \cite{Gerlach69}
\begin{equation}
<g_2,\phi_2,S_2|g_1,\phi_1,S_1> \approx
f(g_2,\phi_2,S_2;g_1,\phi_1,S_1)
e^{\frac{i}{\hbar}I_{cl}[g_2,\phi_2,S_2;g_1,\phi_1,S_1]} ,
\label{4.1}
\end{equation}
where the function outside of the exponential is a slowly varying
function and $I_{cl}$ is the action for a classical 4-geometry. 
Substituting (\ref{4.1}) into (\ref{3.1}) gives
\begin{eqnarray}
<g_2,\phi_2,S_2|\psi> & = & \int f(g_2,\phi_2,S_2;g_1,\phi_1,S_1)
e^{\frac{i}{\hbar} I_{cl}[g_2,\phi_2,S_2;g_1,\phi_1,S_1] }
\nonumber \\
 & & <g_1,\phi_1,S_1|\psi> D(g_1) D(\phi_1) . \label{4.2}
\end{eqnarray}

	Each value of the integrand in (\ref{4.2}) corresponds to
one classical 4-geometry.  As in (\ref{2.3}), there will be an
infinite number of classical 4-geometries that contribute to each
value of the final wave function.  Here, however, we do not have
only one single integration, but an infinite number of
integrations, because the integration is carried out over all possible 3-geometries and all
matter fields on the initial surface.

	In the simple example in Section 2, there were two cases to
consider for the single integration being carried out.  In the
first case, the classical action was not a sharply peaked
function.  In the second case, the classical action was a sharply
peaked function so that a saddlepoint approximation could be
applied to the integration.  Following that strategy, we would
need to consider those two cases for each of the infinite number
of integrations in (\ref{4.2}).  

	Here, however, we consider only the case where $I_{cl}$ is a
sharply peaked function of $g_{1}$ and matter fields $\phi_{1}$
for each of the infinite number of integrations in (\ref{4.2}). 
We consider this case in the following section.
\section{Saddlepoint approximation for the integral over initial
states}
	We consider the case here where $I_{cl}$ is a sharply peaked
function of $g_{1}$ and matter fields $\phi_{1}$ for each of the
infinite number of integrations in (\ref{4.2}).  In that case, we
can formally make the saddlepoint approximation for each of the
integrations in (\ref{4.2}).  In analogy with (\ref{2.4}), we
have the saddlepoint condition
\begin{equation}
\left.\frac{\partial}{\partial g_1}
I_{cl}[g_2,\phi_2,S_2;g_1,\phi_1,S_1] \right|_{g_1=g_{sp}} = 0
\label{5.1}
\end{equation}
and
\begin{equation}
\left.\frac{\partial}{\partial \phi_1}
I_{cl}[g_2,\phi_2,S_2;g_1,\phi_1,S_1] \right|_{\phi_1=\phi_{sp}}
= 0 , \label{5.2}
\end{equation}
where the derivatives in (\ref{5.1}) and (\ref{5.2}) are with
respect to each parameter that defines the 3-geometry $g_1$ and
matter fields $\phi_1$.  We consider the case where there is only
one solution to the saddlepoint conditions (\ref{5.1}) and
(\ref{5.2}).  In that case, (\ref{5.1}) selects a single
classical 4-geometry.  However, there will be a range of
classical 4-geometries in the neighborhood that contribute
significantly to the integral in (\ref{4.2}).  These are
determined by (e.g. \cite{CourantHilbert53})
\begin{equation}
\left| I_{cl}[g_2,\phi_2,S_2;g_1,\phi_1,S_1] -
I_{cl}[g_2,\phi_2,S_2;g_{sp},\phi_{sp},S_1] \right| < \hbar .
\label{5.3}
\end{equation}
We can formally write the saddlepoint approximation to the
integration in (\ref{4.2}) as
{\samepage
\begin{eqnarray}
<g_2,\phi_2,S_2|\psi> & = &
f(g_2,\phi_2,S_2;g_{sp},\phi_{sp},S_1)
<g_{sp},\phi_{sp},S_1|\psi> \nonumber \\
 & & f_1(g_2,\phi_2,S_2;g_{sp},\phi_{sp},S_1) e^{\frac{i}{\hbar}
I_{cl}[g_2,\phi_2,S_2;g_{sp},\phi_{sp},S_1] } \label{5.4}
\end{eqnarray}
}
where classical 4-geometries that contribute significantly to
(\ref{5.4}) (through the function $f_1$) lie within a narrow
range specified by (\ref{5.3}).

	Equation (\ref{5.1}) requires that the momentum canonical to
the initial 3-geometry for the classical 4-geometry at the
saddlepoint be zero.  That is
\begin{equation}
\left.\pi^{ij} \right|_{g_1=g_{sp}} = 0. \label{5.5}
\end{equation}
(The extrinsic curvature on $S_{1}$ will therefore also be zero
at the saddlepoint.)  However, there will be a range of initial
canonical momenta and a range of initial 3-geometries
corresponding to the range of classical 4-geometries that satisfy
(\ref{5.3}), so that the uncertainty relations between initial
3-geometries and their canonical momenta are satisfied.

	Whether there is a narrow or broad range of classical
4-geometries that satisfy (\ref{5.3}) depends on the second
derivative of the action with respect to the initial 3-geometry.  

	We can take the action to be
\begin{equation}
I = \int (-g^{(4)})^{1/2} (L_{geom} + L_{matter}) d^4x +
\frac{1}{8\pi} \int (g^{(3)})^{1/2} K d^3x , \label{6.1}
\end{equation}
where \cite{York72,Hawking79} show the importance of the surface
term.  \cite{Hawking79} also points out a potential problem in
that the action can be changed by conformal transformations, but
suggests a solution.  
\begin{equation}
K = g^{(3)ij} K_{ij} \label{6.2}
\end{equation}
is the trace of the extrinsic curvature.  Although the extrinsic
curvature is zero on $S_{1}$ at the saddlepoint, it will be
nonzero in a region around the saddlepoint.  The extrinsic
curvature is given by
\begin{equation}
K_{ij} = -\frac{1}{2} \frac{\partial g_{ij}^{(3)}}{\partial t} ,
\label{6.3}
\end{equation}
where $g_{ij}^{(3)}$ is the 3-metric.  In this example, we take
the Lagrangian for the geometry as
\begin{equation}
L_{geom} = \frac{R}{16\pi} , \label{6.4}
\end{equation}
where R is the scalar curvature, but we realize that a different
Lagrangian might eventually be shown to be more appropriate in a
correct theory of quantum gravity.

\section{Perfect fluid models}

	For a perfect fluid, the energy momentum tensor is
\begin{equation}
T^{\mu\nu} = (\rho + p) u^{\mu} u^{\nu} + p g^{\mu\nu} ,
\label{6.5}
\end{equation}
where p is the pressure, $\rho$ is the density, and u is the
4-velocity.  For solutions to Einstein's field equations for a
perfect fluid, (\ref{6.4}) becomes
\begin{equation}
L_{geom} = \frac{1}{2} \rho - \frac{3}{2} p , \label{6.6}
\end{equation}
and we can take the Lagrangian for the matter as
\cite{SchutzSorkin77}
\begin{equation}
L_{matter} = \rho \label{6.7a} 
\end{equation}
Thus, the classical action for perfect fluids is
\begin{equation}
I_{cl} = \frac{3}{2} \int (-g^{(4)})^{1/2} (\rho-p) d^4x -
\frac{1}{16\pi} \int (g^{(3)})^{1/2} g^{(3)ij} \frac{\partial
g^{(3)}_{ij}}{\partial t} d^3x .  \label{6.8}
\end{equation}
We can take 
\begin{equation}
p = (\gamma - 1) \rho \label{6.9}
\end{equation}
for the equation of state, where $1 \leq \gamma < 2$.  Then
(\ref{6.8}) becomes
\begin{equation}
I_{cl} = \frac{3}{2} \int (-g^{(4)})^{1/2} (2-\gamma) \rho d^4x -
\frac{1}{16\pi} \int (g^{(3)})^{1/2} g^{(3)ij} \frac{\partial
g^{(3)}_{ij}}{\partial t} d^3x . \label{6.10}
\end{equation}

	Equation (\ref{6.10}) diverges for a spatially open
universe.  The significance of that might be that only spatially
closed universes make sense.  On the other hand, it might be that
the calculation of the action for the correct theory of quantum
gravity will give a finite value for the action, even for a
spatially open universe, but here, we shall restrict our
calculation to the case of a spatially closed universe.

\section{Spatially homogeneous spacetimes} 

	The integration in (\ref{4.2}) is an integration over
functions $g_{1}$ and $\phi_{1}$ defined on $S_{1}$.  In that
sense, it is similar to a path integral.  For example, there are
six independent functions that define $g_{1}$.  As in the
integration for a path integral, there are approximations that
can be made to reduce the number of integrations that must be
performed.  

	Here, we want to consider matter distributions similar to
that observed, at least for the large scale in our universe. 
Thus, we want to restrict the integration in (\ref{4.2}) to
classical spatially homogeneous 4-geometries that have a
homogeneous matter distribution in calculating the classical
action in the exponential.  The integration in (\ref{4.2}) would
then be over the 3-geometries that form the boundary of those
4-geometries on $S_{1}$.
 
	As an example, we shall use Einstein's General Relativity
for the classical 4-geometries, but the same calculations could
be done for other classical gravitational theories, in case it
turns out that General Relativity is not the correct theory of
gravity.  Thus, we want to consider the integration in
(\ref{4.2}) in which the classical 4-geometries used to calculate
the action in the exponential are restricted to Bianchi
cosmologies.  

	The appropriate calculation would be to consider the most
general Bianchi model, with all of the parameters that describe
that model, and carry out the integration over all of those
parameters.  We notice that the Bianchi parameters (which are
time independent) define the initial three geometry, and
therefore are valid integration variables in (\ref{4.2}).  On the
other hand, if it is suspected that the saddlepoint for the
integration will correspond to the Friedmann-Robertson-Walker
(FRW) model, then one can restrict consideration to only those
Bianchi models that include the FRW model as a special case, and
consider integration in (\ref{4.2}) for only one Bianchi
parameter at a time, holding the others fixed at the FRW value. 
Here, we do that for only one of the Bianchi models for
illustration.

	In choosing which Bianchi model to use, we would like one
that has a parameter that can be varied continuously to give the
FRW model.  In addition, we would like to choose a parameter that
represents rotation of inertial frames relative to the matter
distribution.  In that way, we could directly test the ability of
quantum selection to implement Mach's ideas about inertia.  

	So far, I have not been able to find a completely
satisfactory example.  Although the Bianchi IX cosmology is often
used to represent anisotropy, it seems inappropriate for the
present case because it is only a superposition of gravitational
waves on a Friedman-Robertson-Walker background.  

	The Bianchi $VI_{h}$ model seems to be a better homogeneous
model that has a parameter that represents an angular velocity of
inertial frames relative to matter, and setting that parameter to
zero seems to give the FRW metric.  However, there  seem to be
some difficulties with the Bianchi $VI_{h}$ model being able to
change continuously into the FRW model, and also a possible
problem with the topology.  Until I find a better example,
however, I shall use this one.

	We use the solution for the Bianchi $VI_{h}$ model from
\cite{EllisMacCallum69} with $h=-1/9$.  This cosmological model
is relevant here because it has a relative rotation of inertial
frames with respect to the matter.  Specifically,
\begin{equation}
\Omega(t) = \frac{b}{Y^2(t)Z(t)} \label{omega}
\end{equation}
is the angular velocity in the rest frame of an observer moving
with the fluid, of a set of Fermi-propagated axes with respect to
a particular inertial triad.  The parameter $b$ is an arbitrary
constant of the model, and is zero if and only if there is no
rotation of inertial frames relative to matter.  Thus, we are
interested to know the dependence of the classical action on $b$.  

	After some algebra, we have
\begin{equation}
I_{cl} = \frac{3\pi^2}{4a_0} \int_{t_0}^t \frac{Y(t)Z(t)}{X(t)}
dt , \label{6.11}
\end{equation}
where the spatial part of the 4-volume integration has already
been carried out, $a_{0}$ is a parameter of the model, $t_0$
corresponds to the surface $S_1$ in (\ref{3.1}) and (\ref{4.2})
and is enough larger than the Planck time $T^{*}$ that the
semiclassical approximation is valid, the upper limit in
(\ref{6.11}) corresponds to the surface $S_2$ in (\ref{3.1}) and
(\ref{4.2}), and $X(t)$, $Y(t)$, and $Z(t)$ are functions of the
model that must be determined by differential equations given by
\cite{EllisMacCallum69}.  As expected, the surface term in
(\ref{6.10}) has canceled.  

	If we define
\begin{equation}
r^3(t) = X(t)Y(t)Z(t) \left(\frac{-3k}{3a_0^2+q_0^2}\right)^{3/2}
\label{6.12}
\end{equation}
and
\begin{equation}
1+\alpha(t) = \frac{Y(t)^{2/3}Z(t)^{2/3}}{X(t)^{4/3}}
\label{6.13}
\end{equation}
then the classical action in (\ref{6.11}) becomes
\begin{equation}
I_{cl} =
\frac{3\pi^2}{4}\left(\frac{3+q_0^2/a_0^2}{-3k}\right)^{1/2}
\int_{r_0}^r \frac{(1+\alpha)r}{\dot{r}} dr , \label{i2}
\end{equation}
where $\dot{r}=dr/dt$, $r_0$ is enough larger than the Planck
length $L^*$ that quantum effects can be neglected, and $k=+1$
for a closed universe.  

	For the $h = -1/9$ case, we have
\begin{equation}
q_0=-3a_0 \label{q1ninth}
\end{equation}
if and only if $b \neq 0$.  Substituting (\ref{q1ninth}) into
(\ref{i2}) gives
\begin{equation}
I_{cl} = \frac{3\pi^2}{2}\left(\frac{-1}{k}\right)^{1/2}
\int_{r_0}^r \frac{(1+\alpha)r}{\dot{r}} dr , \label{i3}
\end{equation}

	The form of the equation of state in (\ref{6.9}) allows one
of the differential equations for the model to be integrated in
closed form to give
\begin{equation}
8\pi\rho = 3r_m^{3\gamma-2}r^{-3\gamma} \label{rho}
\end{equation}
where $r_m$ is a constant of integration that depends on the
amount of matter in the universe and the speed of expansion
relative to the gravitational attraction.  Equation (\ref{rho})
shows that $r_m$ is a measure of the amount of matter in the
universe for a given value of r.  Therefore, we might expect
Machian effects (inertial induction) to increase for larger
values of $r_m$.

	Using (\ref{rho}), we have
\begin{equation}
\dot{r}^2 = \left(\frac{r_m}{r}\right)^{3\gamma-2} - k - k\alpha
-\frac{k}{(2a_0)^6}\frac{b^2}{3(1+\alpha)^2r^4} +
\frac{r^2}{12}\left(\frac{\dot{\alpha}}{1+\alpha}\right)^2.
\label{rsquare3}
\end{equation}
For the isotropic case, only the first two terms on the right
hand side of (\ref{rsquare3}) are nonzero.  $r_m$ is the value of
r where those two terms are equal.  For a closed universe for the
isotropic case, $r_m$ is the maximum value of r.  

	Equation (\ref{rsquare3}) can be written
\begin{equation}
\dot{r} = \sqrt{ \left(\frac{r}{r_m}\right) ^{2-3\gamma} -
k(1+\alpha) - \frac{k}{(2a_0)^6}\frac{b^2}{3(1+\alpha)^2r^4} +
\frac{V^2}{3r^4}}, \label{6.14}
\end{equation}
and the remaining differential equations to solve are
\begin{equation}
\dot{V} = 3k(1+\alpha)r -
\frac{k}{(2a_0)^6}\frac{2b^2}{(1+\alpha)^2r^3} \label{6.15}
\end{equation}
and
\begin{equation}
\frac{\dot{\alpha}}{1+\alpha} = \frac{2V}{r^3}, \label{6.16}
\end{equation}
where (\ref{6.16}) is the definition of $V(t)$, and $r(t)$
represents the size of the universe.  The angular rotation
(\ref{omega}) of an inertial frame relative to local matter is
given by
\begin{equation}
\Omega(t) = \left(\frac{-k}{4a_0^2}\right)^{3/2}
\frac{b}{[1+\alpha(t)]r^3(t)} \label{6.17}
\end{equation}

	It is not possible to solve the differential equations
exactly in closed form, but we can find approximate solutions. 
In the limit as $b$ approaches zero, it is valid to neglect all
but the first term under the radical in (\ref{6.14}).  The
appendix then gives
\begin{eqnarray}
1+\alpha & \approx & (1+\alpha_0) exp\left\{
\frac{2r_0^3}{3\gamma-6}\frac{\dot{\alpha}_0}{1+\alpha_0}
(r^{\frac{3}{2}\gamma-3} - r_0^{\frac{3}{2}\gamma-3})
r_m^{1-\frac{3}{2}\gamma} + \right.
\nonumber \\
 & & 
12k 
\frac{(3\gamma-6)r^{3\gamma-2} -
(6\gamma-4)r_0^{\frac{3}{2}\gamma+1}r^{\frac{3}{2}\gamma-3} +
(3\gamma+2)r_0^{3\gamma-2}}
{(3\gamma+2)(3\gamma-2)(3\gamma-6) r_m^{3\gamma-2} } 
\nonumber \\
 & & 
\left. - \frac{8k}{3}\frac{b^2}{(2a_0)^6} 
\frac{r^{3\gamma-6}-2r_0^{\frac{3}{2}\gamma-3}r^{\frac{3}{2}\gamma-3}
+ r_0^{3\gamma-6}} {(\gamma-2)(3\gamma-6)r_m^{3\gamma-2} }
\right\} 
\nonumber \\
& \approx & (1+\alpha_0) \left\{1 + 
\frac{2r_0^3}{3\gamma-6}\frac{\dot{\alpha}_0}{1+\alpha_0}
(r^{\frac{3}{2}\gamma-3} - r_0^{\frac{3}{2}\gamma-3})
r_m^{1-\frac{3}{2}\gamma} + \right.
\nonumber \\
 & & 
12k 
\frac{(3\gamma-6)r^{3\gamma-2} -
(6\gamma-4)r_0^{\frac{3}{2}\gamma+1}r^{\frac{3}{2}\gamma-3} +
(3\gamma+2)r_0^{3\gamma-2}}
{(3\gamma+2)(3\gamma-2)(3\gamma-6) r_m^{3\gamma-2} } 
\nonumber \\
 & & 
\left. - \frac{8k}{3}\frac{b^2}{(2a_0)^6} 
\frac{r^{3\gamma-6}-2r_0^{\frac{3}{2}\gamma-3}r^{\frac{3}{2}\gamma-3}
+ r_0^{3\gamma-6}} {(\gamma-2)(3\gamma-6)r_m^{3\gamma-2} }
\right\} 
\nonumber \\
& \approx & (1+\alpha_0) \left\{1 + 
\right.
\nonumber \\
 & & 
12k 
\frac{r^{3\gamma-2} }
{(3\gamma+2)(3\gamma-2) r_m^{3\gamma-2} } 
\nonumber \\
 & & 
\left. - \frac{8k}{3}\frac{b^2}{(2a_0)^6} 
\frac{r_0^{3\gamma-6}} {(\gamma-2)(3\gamma-6)r_m^{3\gamma-2} }
\right\} . 
\label{alpha12}
\end{eqnarray}
This solution is valid for $r$ smaller than $r_m$ if $b$ is small
enough.  

	We assume that $\gamma$ changes from $\gamma_1$ to
$\gamma_2$ at $r=r_1$.  To satisfy continuity of $\rho$ at
$r=r_1$, we must have $r_m$ change from $r_{m1}$ to $r_{m2}$ at
$r=r_1$, where
\begin{equation}
\frac{r_{m1}}{r_{m2}} = \left(\frac{r_1}{r_{m2}}\right)^
{\frac{3\gamma_1-3\gamma_2}{3\gamma_1-2}}. \label{rm1}
\end{equation}
Equation (\ref{alpha12}) applies for $r\leq r_1$ with
$\gamma=\gamma_1$ and $r_m=r_{m1}$.  

	For $r>r_1$, (\ref{alpha12}) gives
\begin{eqnarray}
1+\alpha & \approx & (1+\alpha_1) exp\left\{
\frac{2r_1^3}{3\gamma_2-6}\frac{\dot{\alpha}_1}{1+\alpha_1}
(r^{\frac{3}{2}\gamma_2-3} - r_1^{\frac{3}{2}\gamma_2-3})
r_{m2}^{1-\frac{3}{2}\gamma_2} + \right.
\nonumber \\
 & & 
12k 
\frac{(3\gamma_2-6)r^{3\gamma_2-2} -
(6\gamma_2-4)r_1^{\frac{3}{2}\gamma_2+1}r^{\frac{3}{2}\gamma_2-3}
+ (3\gamma_2+2)r_1^{3\gamma_2-2}}
{(3\gamma_2+2)(3\gamma_2-2)(3\gamma_2-6) r_{m2}^{3\gamma_2-2} } 
\nonumber \\
 & & 
\left. - \frac{8k}{3}\frac{b^2}{(2a_0)^6} 
\frac{r^{3\gamma_2-6}-2r_1^{\frac{3}{2}\gamma_2-3}r^{\frac{3}{2}\gamma_2-3}
+ r_1^{3\gamma_2-6}}
{(\gamma_2-2)(3\gamma_2-6)r_{m2}^{3\gamma_2-2} } \right\} 
\nonumber \\
& \approx & (1+\alpha_1) \left\{1 + 
\frac{2r_1^3}{3\gamma_2-6}\frac{\dot{\alpha}_1}{1+\alpha_1}
(r^{\frac{3}{2}\gamma_2-3} - r_1^{\frac{3}{2}\gamma_2-3})
r_{m2}^{1-\frac{3}{2}\gamma_2} + \right.
\nonumber \\
 & & 
12k 
\frac{(3\gamma_2-6)r^{3\gamma_2-2} -
(6\gamma_2-4)r_1^{\frac{3}{2}\gamma_2+1}r^{\frac{3}{2}\gamma_2-3}
+ (3\gamma_2+2)r_1^{3\gamma_2-2}}
{(3\gamma_2+2)(3\gamma_2-2)(3\gamma_2-6) r_{m2}^{3\gamma_2-2} } 
\nonumber \\
 & & 
\left. - \frac{8k}{3}\frac{b^2}{(2a_0)^6} 
\frac{r^{3\gamma_2-6}-2r_1^{\frac{3}{2}\gamma_2-3}r^{\frac{3}{2}\gamma_2-3}
+ r_1^{3\gamma_2-6}}
{(\gamma_2-2)(3\gamma_2-6)r_{m2}^{3\gamma_2-2} } \right\} 
\nonumber \\
& \approx & (1+\alpha_1) , \label{alpha12a}
\end{eqnarray}
where continuity of $\alpha$ and $\dot{\alpha}$ at $r=r_1$
requires that
\begin{eqnarray}
1+\alpha_1 & \approx & (1+\alpha_0) exp\left\{
\frac{2r_0^3}{3\gamma_1-6}\frac{\dot{\alpha}_0}{1+\alpha_0}
(r_1^{\frac{3}{2}\gamma_1-3} - r_0^{\frac{3}{2}\gamma_1-3})
r_{m1}^{1-\frac{3}{2}\gamma_1} + \right.
\nonumber \\
 & & 
12k 
\frac{(3\gamma_1-6)r_1^{3\gamma_1-2} -
(6\gamma_1-4)r_0^{\frac{3}{2}\gamma_1+1}r_1^{\frac{3}{2}\gamma_1-3}
+ (3\gamma_1+2)r_0^{3\gamma_1-2}}
{(3\gamma_1+2)(3\gamma_1-2)(3\gamma_1-6) r_{m1}^{3\gamma_1-2} } 
\nonumber \\
 & & 
\left. - \frac{8k}{3}\frac{b^2}{(2a_0)^6} 
\frac{r_1^{3\gamma_1-6}-2r_0^{\frac{3}{2}\gamma_1-3}r_1^{\frac{3}{2}\gamma_1-3}
+ r_0^{3\gamma_1-6}}
{(\gamma_1-2)(3\gamma_1-6)r_{m1}^{3\gamma_1-2} } \right\} 
\nonumber \\
& \approx & (1+\alpha_0) \left\{1 + 
\frac{2r_0^3}{3\gamma_1-6}\frac{\dot{\alpha}_0}{1+\alpha_0}
(r_1^{\frac{3}{2}\gamma_1-3} - r_0^{\frac{3}{2}\gamma_1-3})
r_{m1}^{1-\frac{3}{2}\gamma_1} + \right.
\nonumber \\
 & & 
12k 
\frac{(3\gamma_1-6)r_1^{3\gamma_1-2} -
(6\gamma_1-4)r_0^{\frac{3}{2}\gamma_1+1}r_1^{\frac{3}{2}\gamma_1-3}
+ (3\gamma_1+2)r_0^{3\gamma_1-2}}
{(3\gamma_1+2)(3\gamma_1-2)(3\gamma_1-6) r_{m1}^{3\gamma_1-2} } 
\nonumber \\
 & & 
\left. - \frac{8k}{3}\frac{b^2}{(2a_0)^6} 
\frac{r_1^{3\gamma_1-6}-2r_0^{\frac{3}{2}\gamma_1-3}r_1^{\frac{3}{2}\gamma_1-3}
+ r_0^{3\gamma_1-6}}
{(\gamma_1-2)(3\gamma_1-6)r_{m1}^{3\gamma_1-2} } 
\right\}
\nonumber \\
& \approx & (1+\alpha_0) \left\{1 + 
\right.
\nonumber \\
 & & 
12k 
\frac{r_1^{3\gamma_1-2} }
{(3\gamma_1+2)(3\gamma_1-2) r_{m1}^{3\gamma_1-2} } 
\nonumber \\
 & & 
\left. - \frac{8k}{3}\frac{b^2}{(2a_0)^6} 
\frac{r_0^{3\gamma_1-6}}
{(\gamma_1-2)(3\gamma_1-6)r_{m1}^{3\gamma_1-2} } 
\right\} . \label{alpha12b}
\end{eqnarray}

	We then perform the integration in (\ref{i3}).  The result
is
\begin{eqnarray}
\frac{I_{cl}}{\hbar} & = &   \frac{3\pi^2}{2} (-k)^{1/2}
\left(\frac{r_{m2}}{L^*}\right)^2 (1+\alpha_0)\left\{
\frac{2}{3\gamma_1+2} 
\left[\left(\frac{r_1}{r_{m1}}\right)^{\frac{3}{2}\gamma_1+1} -
\left(\frac{r_0}{r_{m1}}\right)^{\frac{3}{2}\gamma_1+1} \right] 
\left(\frac{r_{m1}}{r_{m2}}\right)^2 
\right.  \nonumber \\
 & & + 
\frac{24k
\left[\left(\frac{r_1}{r_{m1}}\right)^{\frac{9}{2}\gamma_1-1} -
\left(\frac{r_0}{r_{m1}}\right)^{\frac{9}{2}\gamma_1-1} \right]
} {(3\gamma_1+2)(3\gamma_1-2)(9\gamma_1-2)} 
\left(\frac{r_{m1}}{r_{m2}}\right)^2 
\nonumber \\
 & & -  
\frac{8k}{3} \frac{b^2r_{m1}^{-4}}{(2a_0)^6} \frac{
\left[\left(\frac{r_1}{r_{m1}}\right)^{\frac{3}{2}\gamma_1+1} -
\left(\frac{r_0}{r_{m1}}\right)^{\frac{3}{2}\gamma_1+1} \right]
 } {(\gamma_1-2)^2 3(3\gamma_1+2)} 
2\left(\frac{r_0}{r_{m1}}\right)^{3\gamma_1-6} 
\left(\frac{r_{m1}}{r_{m2}}\right)^2 
\nonumber \\
 & & + 
\left[1+12k\frac{r_1^{3\gamma_1-2}r_{m1}^{2-3\gamma_1}}
{(3\gamma_1+2)(3\gamma_1-2)} 
-\frac{8k}{3} \frac{b^2r_{m1}^{-4}}{(2a_0)^6} 
\frac{r_0^{3\gamma_1-6}r_{m1}^{6-3\gamma_1}}{(\gamma_1-2)(3\gamma_1-6)}
\right] 
 \nonumber \\
 & & \left. \frac{2}{3\gamma_2+2} 
\left[\left(\frac{r}{r_{m2}}\right)^{\frac{3}{2}\gamma_2+1} -
\left(\frac{r_1}{r_{m2}}\right)^{\frac{3}{2}\gamma_2+1} \right] 
\right\} . \label{i4}
\end{eqnarray}
Equation (\ref{i4}) neglects all but the first term under the
radical in (\ref{6.14}).  In making the calculation, I actually
included the other terms under the radical to first order, but
then determined after the integration that they could be
neglected.

	Neglecting some small terms gives
\begin{eqnarray}
\frac{I_{cl}}{\hbar} & = &   \frac{3\pi^2}{2} (-k)^{1/2}
\left(\frac{r_{m2}}{L^*}\right)^2 (1+\alpha_0)
\frac{2}{3\gamma_2+2} 
\left(\frac{r}{r_{m2}}\right)^{\frac{3}{2}\gamma_2+1} 
\nonumber \\
 & & 
\left[1
-\frac{8k}{3} \frac{b^2r_{m1}^{-4}}{(2a_0)^6} 
\frac{r_0^{3\gamma_1-6}r_{m1}^{6-3\gamma_1}}{(\gamma_1-2)(3\gamma_1-6)}
\right] 
. \label{i4a}
\end{eqnarray}
Substituting (\ref{rm1}) into (\ref{i4a}) gives
\begin{eqnarray}
\frac{I_{cl}}{\hbar} & = &   \frac{3\pi^2}{2} (-k)^{1/2}
\left(\frac{r_{m2}}{L^*}\right)^2 (1+\alpha_0)
\frac{2}{3\gamma_2+2} 
\left(\frac{r}{r_{m2}}\right)^{\frac{3}{2}\gamma_2+1} 
\nonumber \\
 & & 
\left[1
-\frac{8k}{3}
\left(\frac{r_1}{r_{m2}}\right)^{3\gamma_2-3\gamma_1}
\frac{b^2r_{m2}^{-4}}{(2a_0)^6} 
\frac{r_0^{3\gamma_1-6}r_{m2}^{6-3\gamma_1}}{(\gamma_1-2)(3\gamma_1-6)}
\right] 
. \label{i4b}
\end{eqnarray}

	To get a rough estimate, we take 
\begin{equation}
\gamma_1 = 4/3 \label{gamma1}
\end{equation}
to represent a relativistic early universe and 
\begin{equation}
\gamma_2 = 1 \label{gamma2}
\end{equation}
to represent a matter-dominated late universe.  Substituting
(\ref{gamma1}) and (\ref{gamma2}) into (\ref{i4}) gives
\begin{eqnarray}
\frac{I_{cl}}{\hbar} & = &   \frac{3\pi^2}{2} (-k)^{1/2}
\left(\frac{r_{m2}}{L^*}\right)^2 (1+\alpha_0) \left\{
\frac{1}{3} 
\left[\left(\frac{r_1}{r}\right)^{3}\left(\frac{r}{r_{m1}}\right)^{3}
- \left(\frac{r_0}{r_{m1}}\right)^{3} \right] 
\left(\frac {r_{m1}} {r_{m2}} \right)^2  
\right. \nonumber \\
 & & + \frac{2}{5} 
\left[\left(\frac{r}{r_{m2}}\right)^{\frac{5}{2}} -
\left(\frac{r_1}{r}\right)^{5/2}\left(\frac{r}{r_{m2}}\right)^{\frac{5}{2}}
\right] \left[1+\left(\frac{r_1}{r}\right)^{2}k\left(\frac {r}
{r_{m1}} \right)^2  \right]
 \nonumber \\
 & & + 
\frac{k} {5} 
\left[\left(\frac{r_1}{r}\right)^{5}\left(\frac{r}{r_{m1}}\right)^{5}
- \left(\frac{r_0}{r_{m1}}\right)^{5} \right]
\left(\frac {r_{m1}} {r_{m2}} \right)^2  
\nonumber \\
 & & -  
\frac{2kb^2} {(2a_0)^6 r_{m1}^2 r_0^2 } 
\nonumber \\
 & & 
\left[\frac{1}{3}
\left(\frac {r_{m1}} {r_{m2}} \right)^2  
\left(\left(\frac{r_1}{r}\right)^{3}
\left(\frac{r}{r_{m1}}\right)^3 - 
\left(\frac{r_0}{r_{m1}}\right)^3 
\right) 
\right. 
\nonumber \\
 & & 
+
\left.
\left.
\frac{2}{5}
\left(\frac{r}{r_{m2}}\right)^{5/2} 
\left(1-\left(\frac{r_1}{r}\right)^{5/2}\right) 
\right]
\right\} . \label{i5}
\end{eqnarray}
Neglecting some small terms, letting $k=+1$ for a closed
universe, and using (\ref{rm1}) gives
\begin{eqnarray}
\frac{I_{cl}}{\hbar} & = &   \frac{3i\pi^2}{2}
\left(\frac{r_{m2}}{L^*}\right)^2 
(1+\alpha_0) 
\left\{
\frac{1}{3} \left(\frac{r_{1}}{r}\right)^{3}
 \left(\frac{r}{r_{m2}}\right)^{5/2} 
 \left(\frac{r}{r_{1}}\right)^{1/2} 
\right. \nonumber \\
 & & + 
\frac{2}{5} 
\left[1 + 
\left(\frac{r_{1}}{r}\right)^{2} 
\left(\frac{r}{r_{m2}}\right)^2 
\frac{r_{m2}}{r_1}\right]
\left(\frac{r}{r_{m2}}\right)^{\frac{5}{2}} 
 + \frac {1} {5} 
\left(\frac{r_{1}}{r}\right)^{5} 
 \left(\frac{r}{r_{m2}}\right)^{7/2} 
 \left(\frac{r}{r_{1}}\right)^{3/2} 
\nonumber \\
 & & - \left.
\frac{2}{5} \frac {2b^2} {(2a_0)^6r_{m2}^2r_0^2} 
\left(\frac{r}{r_{1}}\right)
\left(\frac{r}{r_{m2}}\right)^{3/2} 
 \right\}
\nonumber \\
 & \approx &
 \frac{3i\pi^2}{2} \left(\frac{r_{m2}}{L^*}\right)^2 
(1+\alpha_0) 
\nonumber \\
& & \left\{
\frac{2}{5} 
\left(\frac{r}{r_{m2}}\right)^{\frac{5}{2}} 
- \frac{2}{5} \frac {2b^2} {(2a_0)^6r_{m2}^2r_0^2} 
\left(\frac{r}{r_{1}}\right)
\left(\frac{r}{r_{m2}}\right)^{3/2} 
 \right\} . \label{6.19}
\end{eqnarray}

	Because the parameter $b$ is an initial value for the
cosmology, it is one of the variables of integration in
(\ref{4.2}).  In making the saddlepoint approximation for that
integration, we need to locate the saddlepoint (that is, the
value of $b$ that makes the action in (\ref{i4a}), (\ref{i4b}),
or (\ref{6.19})  stationary.  We see that the action is
stationary with respect to variation of $b$ at the isotropic case
of $b=0$, as expected.  The range of values of $b$ that
contribute significantly to the integral in (\ref{4.2}) is given
by (\ref{5.3}).  That is  
\begin{equation}
\left|\frac{I_{cl}(b)}{\hbar} - \frac{I_{cl}(b=0)}{\hbar}\right| 
< 1. \label{6.20}
\end{equation}
Thus, substituting (\ref{i4b}) into (\ref{6.20}) gives
\begin{eqnarray}
&  &   \frac{3\pi^2}{2}  \left(\frac{r_{m2}}{L^*}\right)^2
\frac{(1+\alpha_0) b^2 } {(2a_0)^6 r_{m2}^4} 
\left(\frac{r}{r_{m2}}\right)^{\frac{9}{2}\gamma_2-3\gamma_1+1}
 \nonumber \\
 & &  
\frac{16}{(\gamma_1-2)^2 9(3\gamma_2+2)} 
 \left(\frac{r}{r_{1}}\right)^{3(\gamma_1-\gamma_2)} 
\left(\frac{r_{m2}}{r_0}\right)^{6-3\gamma_1}
< 1. \label{6.21}
\end{eqnarray}
The approximations made so far are valid whenever (\ref{6.21})
holds.

	Substituting (\ref{gamma1}) and (\ref{gamma2}) into
(\ref{6.21}) gives
\begin{equation}
\frac{6\pi^2}{5} (1+\alpha_0) \left(\frac{r}{r_1}\right)
\left(\frac {r} {r_{m2}} \right)^{\frac{3}{2}} 
\left(\frac{L^*}{r_0}\right)^2
\frac{b^2}{(2a_0)^6L^{*4}} 
< 1. \label{6.22}
\end{equation}
This gives
\begin{equation}
b < \frac{\sqrt{5}(2a_0)^3 }{\pi\sqrt{6}}
\left(\frac{r_1}{r}\right)^{1/2} 
\frac{L^{*2}}{(1+\alpha_0)^{1/2}} 
\left(\frac {r_{m2}} {r} \right)^{\frac{3}{4}} 
\left(\frac{r_0}{L^*}\right)
. \label{6.23}
\end{equation}
Thus, from (\ref{6.17}), the rotation rate of inertial frames is
\begin{equation}
\left|\Omega(t)\right| < \frac{\sqrt{5}}{\pi\sqrt{6}}
\left(\frac{L^*}{r_{m2}}\right)^2
\frac{1}{(1+\alpha_0)^{1/2}[1+\alpha(t)]r_{m2}} 
\left(\frac{r_{m2}}{r(t)}\right)^{\frac{15}{4}} 
\left(\frac{r_1}{r}\right)^{1/2}
\left(\frac{r_0}{L^*}\right)
. \label{6.24}
\end{equation}

	If we now take the Planck length $L^*$ to be
$1.6\times10^-33$ cm, use the Hubble distance of
$1.7\times10^{28}$ cm for $r_{m2}$, a tenth of that for $r$,
neglect $\alpha$ and $\alpha_0$ compared to 1, and take
\begin{equation}
r_1 = \frac{r}{100} \label{r1}
\end{equation}
as an estimate that the universe changed from radiation-dominated
to matter-dominated when the universe was about one-hundredth of
its present size \cite[Section 15.3, p. 481]{Weinberg72}, then we
get 
\begin{equation}
|\Omega| < 1.6\times10^{-130} \mbox{ radians per year,}
\label{omega2}
\end{equation}
which is much less than the bound set by experiment of $10^{-14}$
to $7\times 10^{-17}$ radians per year if the universe is
spatially closed \cite{Hawking69}.  

	The rotation rate in (\ref{6.24}) and (\ref{omega2}) is so
small because the Planck length is so much smaller than the
Hubble distance.  

	That the small value of allowed rotation rate depends mostly
on the universe being much larger than a Planck length rather
than on details of the model suggests that the result has some
generality.

	We notice also, that the selection criterion in (\ref{6.20})
is so sharp that the initial wave function in the integration in
(\ref{4.2}) would have to be very sharply peaked to overcome it.  

\section{Discussion}
	We see that considerations of quantum cosmology show how a
range of classical cosmologies can be selected that contribute
significantly to the wave function in the final state.  The
effect enters through the action.  Using semiclassical
calculations gives results that should not depend on particular
details of the theory of quantum gravity.  

	For our universe (which is much larger than the Planck
length) the selection is very sharp.  The initial wave function
over 3-geometries would have to be extremely sharp (not a
probable occurrence) to dominate over the effect of the action.

	The selection process seems to occur very soon in the
development of a cosmology.  That is, for a broad wave function
over 3-geometries in the initial state, the wave function becomes
sharply peaked after the universe has become a few orders of
magnitude larger than the Planck length.

	A different choice than (\ref{6.7a}) \cite{Schutz76} is
\begin{equation}
L_{matter} = p . \label{6.7b} 
\end{equation}
The choice in (\ref{6.7b}) gives a third of (\ref{6.8}) for the
total action.  A correct theory of quantum gravity will determine
which (if either) of these two choices is correct, but for this
illustration, a factor of three in the action makes little
difference.  

	It appears likely now, however, that there is not enough
matter to keep our universe from expanding forever (e.g.
\cite{ColesEllis94}).  To accommodate that with a spatially
closed universe within General Relativity would require a
positive cosmological constant (e.g. \cite{Hawking98}).  I shall
try to include the cosmological constant in a future calculation.

\section{Acknowledgment} 
	I would like to thank Douglas Gough for first bringing the
paper by \cite{Sciama53} to my attention in 1967.

\appendix
\section{Bianchi $VI_h$ Models}
	We start with  Equations (6.7) of Ellis and MacCallum
(1969).  For the case of zero cosmological constant, $\Lambda$,
these can be written as
\begin{equation}
\frac{\dot{\rho}}{\rho+p} = - \frac{3\dot{R}}{R} \label{ellisrho}
\end{equation}
\begin{equation}
4\pi(\rho-p) = R^{-3}\left(R^{3}\frac{\dot{X}}{X}\right)^{.} -
\frac{2(a_{0}^{2} + q_{0}^{2})}{X^{2}} +
\frac{2b^{2}}{Y^{4}Z^{2}} \label{ellisX}
\end{equation}
\begin{equation}
4\pi(\rho-p) = R^{-3}\left(R^{3}\frac{\dot{Y}}{Y}\right)^{.} -
\frac{2(a_{0}^{2} + a_{0}q_{0})}{X^{2}} -
\frac{2b^{2}}{Y^{4}Z^{2}} \label{ellisY}
\end{equation}
\begin{equation}
4\pi(\rho-p) = R^{-3}\left(R^{3}\frac{\dot{Z}}{Z}\right)^{.} -
\frac{2(a_{0}^{2} - a_{0}q_{0})}{X^{2}} \label{ellisZ}
\end{equation}
\begin{equation}
8\pi\rho + \frac{3a_{0}^{2} + q_{0}^{2}}{X^{2}} =
\frac{1}{2}\left[9\left(\frac{\dot{R}}{R}\right)^{2}
-\left(\frac{\dot{X}}{X}\right)^{2}
-\left(\frac{\dot{Y}}{Y}\right)^{2}
-\left(\frac{\dot{Z}}{Z}\right)^{2}\right]
+\frac{2b^{2}}{Y^{4}Z^{2}}, \label{ellisR}
\end{equation}
where

\begin{equation}
R(t)^3 = X(t)Y(t)Z(t), \label{Rdef}
\end{equation}

b, $a_0$, and $q_0$ are constants that are parameters of the
model, and the variation of $\rho$, $X$, $Y$, and $Z$ with time
is determined by (\ref{ellisrho}) through (\ref{ellisR}).  

	According to Ellis and MacCallum (1969), (\ref{ellisR}) is a
first integral of the other equations.  If I understand that
correctly, then taking the derivative of (\ref{ellisR}) should be
a combination of the other equations.  When I try that, the
result is close, but not quite correct.  I have not been able to
find a similar equation that does work, although I have found one
that works for the special case ($q_0+3a_0=0$ and $XY=Z^2$) that
I use later.  This is
\begin{equation}
8\pi\rho + \frac{3a_{0}^{2} + q_{0}^{2}}{X^{2}} =
\frac{1}{2}\left[9\left(\frac{\dot{R}}{R}\right)^{2}
-\left(\frac{\dot{X}}{X}\right)^{2}
-\left(\frac{\dot{Y}}{Y}\right)^{2}
-\left(\frac{\dot{Z}}{Z}\right)^{2}\right]
-\frac{b^{2}}{Y^{4}Z^{2}}. \label{jonesR}
\end{equation}

	We can add (\ref{ellisX}), (\ref{ellisY}), and
(\ref{ellisZ}) with coefficients $A$, $B$, and $C$ to give
\begin{equation}
(A+B+C)\left[4\pi(\rho-p)+\frac{2a_0^2}{X^2}\right] =
R^{-3}\left(R^{3}\frac{\dot{U}}{U}\right)^{.}
-(B-C)\frac{2a_0q_0}{X^2} -2A\frac{q_0^2}{X^2}
+(A-B)\frac{2b^2}{Y^4Z^2}, \label{ABC}
\end{equation}
where
\begin{equation}
U=X^AY^BZ^C. \label{UdefABC}
\end{equation}
We can take special cases of $A$, $B$, and $C$.  Taking
$A=B=C=1/3$ in (\ref{ABC}) gives
\begin{equation}
4\pi(\rho-p) =
R^{-3}(R^{2}\dot{R})^{.}-\frac{2}{3}\frac{3a_0^2+q_0^2}{X^2},
\label{R1}
\end{equation}
where I have used (\ref{Rdef}).  For another special case, we
take $A=B=1$ and $C=-2$ in (\ref{ABC}) to give
\begin{equation}
R^{-3}\left(R^{3}\frac{\dot{U}}{U}\right)^{.} =
-2q_0\frac{3a_0+q_0}{X^2}, \label{U1}
\end{equation}
where
\begin{equation}
U = \frac{XY}{Z^2}. \label{Udef}
\end{equation}
For a third special case, we take $A=-4/3$ and $B=C=2/3$ in
(\ref{ABC}) to give
\begin{equation}
R^{-3}\left(R^{3}\frac{\dot{\alpha}}{1+\alpha}\right)^{.} =
\frac{4b^2}{Y^4Z^2} -\frac{8q_0^2}{3X^2}, \label{alpha1}
\end{equation}
where
\begin{equation}
1+\alpha = \frac{Y^{2/3}Z^{2/3}}{X^{4/3}}. \label{alphadef}
\end{equation}
We can use (\ref{Rdef}), (\ref{Udef}), and (\ref{alphadef}) to
determine $X$, $Y$, and $Z$ in terms of $R$, $U$, and $\alpha$. 
This gives
\begin{equation}
X = \frac{R}{(1+\alpha)^{1/2}}, \label{Xdef}
\end{equation}

\begin{equation}
Y = U^{1/3}(1+\alpha)^{1/2}R, \label{Ydef}
\end{equation}
and
\begin{equation}
Z = \frac{R}{U^{1/3}}. \label{Zdef}
\end{equation}
Using (\ref{Xdef}), (\ref{Ydef}), and (\ref{Zdef}) in (\ref{R1}),
(\ref{U1}), (\ref{alpha1}), and (\ref{jonesR}) gives
\begin{equation}
R^{-3}\left(R^{2}\dot{R}\right)^{.} = 4\pi(\rho-p) 
+\frac{2(3a_0^2+q_0^2)(1+\alpha)}{3R^2}, \label{R2}
\end{equation}
\begin{equation}
R^{-3}\left(R^{3}\frac{\dot{U}}{U}\right)^{.} =
-2q_0\frac{(3a_0+q_0)(1+\alpha)}{R^2}, \label{U2}
\end{equation}
\begin{equation}
R^{-3}\left(R^{3}\frac{\dot{\alpha}}{1+\alpha}\right)^{.} =
\frac{4b^2}{(1+\alpha)^2U^{2/3}R^6} -
\frac{8q_0^2(1+\alpha)}{3R^2}, \label{alpha2}
\end{equation}
and
\begin{eqnarray}
\left(\frac{\dot{R}}{R}\right)^2 & = &
\frac{1}{3}\left[\left(\frac{1}{2}\frac{\dot{\alpha}}{1+\alpha}\right)^2
+ \left(\frac{1}{2}\frac{\dot{\alpha}}{1+\alpha}\right)
\left(\frac{1}{3}\frac{\dot{U}}{U}\right) +
\left(\frac{1}{3}\frac{\dot{U}}{U}\right)^2\right] \nonumber \\
 & & + \frac{8\pi\rho}{3} +\frac{(3a_0^2+q_0^2)(1+\alpha)}{3R^2} 
+\frac{b^2}{3(1+\alpha)^2U^{2/3}R^6}. \label{Rsquare2}
\end{eqnarray}
If we change variables from R to r, where
\begin{equation}
r = \sqrt{\frac{-3k}{3a_0^2+q_0^2}}R,  \label{rdef}
\end{equation}
(and where $k$ is +1 for a closed universe and -1 for an open
universe), then (\ref{R2}) through (\ref{Rsquare2}) become
\begin{equation}
r^{-3}(r^{2}\dot{r})^{.} = 4\pi(\rho-p) 
-\frac{2k(1+\alpha)}{r^2}, \label{r3}
\end{equation}
\begin{equation}
r^{-3}\left(r^{3}\frac{\dot{U}}{U}\right)^{.} =
\frac{6kq_0(3a_0+q_0)(1+\alpha)}{(3a_0^2+q_0^2)r^2}, \label{U3}
\end{equation}
\begin{equation}
r^{-3}\left(r^{3}\frac{\dot{\alpha}}{1+\alpha}\right)^{.} =
\left(\frac{-3k}{3a_0^2+q_0^2}\right)^3\frac{4b^2}{(1+\alpha)^2U^{2/3}r^6}
+ \frac{8kq_0^2(1+\alpha)}{(3a_0^2+q_0^2)r^2}, \label{alpha3}
\end{equation}
and
\begin{eqnarray}
\left(\frac{\dot{r}}{r}\right)^2 & = &
\frac{1}{3}\left[\left(\frac{1}{2}\frac{\dot{\alpha}}{1+\alpha}\right)^2
+ \left(\frac{1}{2}\frac{\dot{\alpha}}{1+\alpha}\right)
\left(\frac{1}{3}\frac{\dot{U}}{U}\right) +
\left(\frac{1}{3}\frac{\dot{U}}{U}\right)^2\right] \nonumber \\
 & & + \frac{8\pi\rho}{3} -\frac{k(1+\alpha)}{r^2} 
+\left(\frac{-3k}{3a_0^2+q_0^2}\right)^3\frac{b^2}{3(1+\alpha)^2U^{2/3}r^6}.
\label{A-rsquare3}
\end{eqnarray}
If we define $W$, $V$, and $K$ by
\begin{equation}
r^2\dot{r} = W, \label{r4}
\end{equation}
\begin{equation}
\frac{\dot{\alpha}}{1+\alpha} = \frac{2V}{r^3} \label{alpha4}
\end{equation}
and
\begin{equation}
\frac{\dot{U}}{U} = \frac{3K}{r^3}, \label{U4}
\end{equation}
then (\ref{r3}) through (\ref{A-rsquare3}) become
\begin{equation}
\dot{W} = 4\pi(\rho-p)r^3  - 2k(1+\alpha)r, \label{W4}
\end{equation}
\begin{equation}
\dot{K} = \frac{2kq_0(3a_0+q_0)(1+\alpha)r}{3a_0^2+q_0^2} ,
\label{K4}
\end{equation}
\begin{equation}
\dot{V} =
\left(\frac{-3k}{3a_0^2+q_0^2}\right)^3\frac{2b^2}{(1+\alpha)^2U^{2/3}r^3}
+ \frac{4kq_0^2(1+\alpha)r}{3a_0^2+q_0^2}, \label{V4}
\end{equation}
\begin{eqnarray}
W^2 & = & \frac{V^2+VK+K^2}{3} + \frac{8\pi\rho r^6}{3} \nonumber
\\
 & & - k(1+\alpha)r^4 +
\left(\frac{-3k}{3a_0^2+q_0^2}\right)^3\frac{b^2}{3(1+\alpha)^2U^{2/3}}.
\label{rsquare4}
\end{eqnarray}
If we take
\begin{equation}
p = (\gamma-1)\rho \label{state}
\end{equation}
for the equation of state, where
\begin{equation}
1 \leq \gamma < 2 \label{gamma}
\end{equation}
is a constant, then (\ref{ellisrho}) can be integrated to give
\begin{equation}
8\pi\rho = 3r_m^{3\gamma-2}r^{-3\gamma} \label{A-rho}
\end{equation}
where $r_m$ is a constant of integration.  Substituting
(\ref{state}) and (\ref{A-rho}) into (\ref{W4}) and
(\ref{rsquare4}) gives
\begin{equation}
\dot{W} = 3(1-\gamma/2)r_m^{3\gamma-2}r^{3-3\gamma} -
2k(1+\alpha)r, \label{r5}
\end{equation}
\begin{eqnarray}
W^2 & = & \frac{V^2+VK+K^2}{3} + r_m^{3\gamma-2}r^{6-3\gamma}
\nonumber \\
 & & - k(1+\alpha)r^4 +
\left(\frac{-3k}{3a_0^2+q_0^2}\right)^3\frac{b^2}{3(1+\alpha)^2U^{2/3}}.
\label{rsquare5}
\end{eqnarray}
Let us now convert to a set of dimensionless variables defined by
\begin{equation}
\tau=t/r_m \label{taudef}
\end{equation}
\begin{equation}
x=r/r_m \label{xdef}
\end{equation}
\begin{equation}
y=V/r_m^2 \label{ydef}
\end{equation}
\begin{equation}
z=W/r_m^2 \label{zdef}
\end{equation}
\begin{equation}
C=K/r_m^2 \label{Cdef}
\end{equation}
\begin{equation}
D=b/r_m^2 \label{Ddef}
\end{equation}
Let us also use $^{\prime}$ for $d/d\tau$.  Then (\ref{r4})
through (\ref{V4}), (\ref{r5}), and (\ref{rsquare5}) become
\begin{equation}
x^2x^{\prime} = z, \label{x6}
\end{equation}
\begin{equation}
\frac{\alpha^{\prime}}{1+\alpha} = \frac{2y}{x^3} \label{alpha6}
\end{equation}
\begin{equation}
\frac{U^{\prime}}{U} = \frac{3C}{x^3}, \label{U6}
\end{equation}
\begin{equation}
C^{\prime} = \frac{2kq_0(3a_0+q_0)(1+\alpha)x}{3a_0^2+q_0^2} ,
\label{K6}
\end{equation}
\begin{equation}
y^{\prime} =
\left(\frac{-3k}{3a_0^2+q_0^2}\right)^3\frac{2D^2}{(1+\alpha)^2U^{2/3}x^3}
+ \frac{4kq_0^2(1+\alpha)x}{3a_0^2+q_0^2}, \label{y6}
\end{equation}
\begin{equation}
z^{\prime} = 3(1-\gamma/2)x^{3-3\gamma} -2k(1+\alpha)x,
\label{z6}
\end{equation}
\begin{equation}
z^2 = x^{6-3\gamma} - k(1+\alpha)x^4 +
\left(\frac{-3k}{3a_0^2+q_0^2}\right)^3\frac{D^2}{3(1+\alpha)^2U^{2/3}}
+ \frac{y^2+yC+C^2}{3}. \label{zsquare6}
\end{equation}

	For the $h = -1/9$ case, we have
\begin{equation}
q_0=-3a_0. \label{A-q1ninth}
\end{equation}
if and only if $b \neq 0$.  However, the $b=0$ case is only a
single point.  When integrating over $b$, the behavior for small
$b$ is more important than exactly at $b=0$.  Therefore, we shall
use (\ref{A-q1ninth}) in any case.  Therefore, from (\ref{K6}),
we have that $C$ is a constant.  

	So both $b$ and $C$ are constants that are determined by
initial conditions.  There are two possibilities.  Either they
are connected by a relation, or they are independent.  There
seems to be no obvious connection between them from the
equations, so we shall assume they are independent unless we
shall find it necessary to make a connection to get a consistent
solution to the equations.  Therefore, we can assume that both
$b$ and $C$ must be integrated over on the initial hypersurface. 
For this demonstration, however, it is sufficient to integrate
over only one initial constant, $b$.  Therefore, we shall fix $C$
at the value we would guess for the FRW case, which is zero. 
Therefore, we take $C$ to be zero.

	That means from (\ref{U6}) that $U$ is constant.  Again, we
guess that $U$ is independent from $b$, and choose the isotropic
value.  In the isotropic case, we would have $X=Y=Z$, and
therefore, from (\ref{Udef}) we have $U=1$ as the isotropic
value.  

	Therefore, from (\ref{y6}) we have 
\begin{equation}
y^{\prime} = 3k(1+\alpha)x -
\frac{k}{(2a_0)^6}\frac{2D^2}{(1+\alpha)^2x^3}, \label{y7}
\end{equation}
and from (\ref{zsquare6}) we have
\begin{equation}
z^2 = x^{6-3\gamma} - k(1+\alpha)x^4 -
\frac{k}{(2a_0)^6}\frac{D^2}{3(1+\alpha)^2} + \frac{y^2}{3}.
\label{zsquare7}
\end{equation}
We notice at this point that if we take the derivative of
(\ref{zsquare7}) and substitute from (\ref{x6}), (\ref{alpha6}),
(\ref{y7}), and (\ref{z6}), that we get an identity, confirming
the consistency of the equations at this point.

	Combining (\ref{x6}) with (\ref{zsquare7}) gives
\begin{equation}
x^{\prime} = \sqrt{x^{2-3\gamma} - k(1+\alpha) -
\frac{k}{(2a_0)^6}\frac{D^2}{3(1+\alpha)^2x^4} +
\frac{y^2}{3x^4}}. \label{x7}
\end{equation}
Combining (\ref{y7}) with (\ref{x7}) gives
\begin{equation}
\frac{dy}{dx} = \frac{3k(1+\alpha)x -
\frac{k}{(2a_0)^6}\frac{2D^2}{(1+\alpha)^2x^3}}
{\sqrt{x^{2-3\gamma} - k(1+\alpha) -
\frac{k}{(2a_0)^6}\frac{D^2}{3(1+\alpha)^2x^4} +
\frac{y^2}{3x^4}}}. \label{dydx7}
\end{equation}

	Both $b$ and $a_0$ are constants that are determined by
initial conditions on the initial hypersurface.  They are either
related or independent.  We assume first that they are
independent.  In that case, the third term under the radical in
(\ref{x7}) and (\ref{dydx7}) will get smaller as $D$ gets
smaller.  We shall assume that we can neglect that term relative
to the first term under the radical for all values of $x$.  We
can check that assumption later.  There is also the possibility
of iterating later by assuming that this term is small instead of
zero.

	We shall also neglect the last term under the radical in
(\ref{x7}) and (\ref{dydx7}).  We can check that approximation
later.  There is also the possibility of iterating by
substituting an approximate solution for $y$ into (\ref{dydx7}). 
We shall see that it it permissible to neglect the fourth term
for all values of $x$ within the integration range if $b$ is
small enough.  

	Neglecting the third and fourth terms under the radical in
(\ref{x7}) and (\ref{dydx7}) gives
\begin{equation}
x^{\prime} = \sqrt{x^{2-3\gamma} - k(1+\alpha)}. \label{x8}
\end{equation}
and
\begin{equation}
\frac{dy}{dx} = \frac{3k(1+\alpha)x -
\frac{k}{(2a_0)^6}\frac{2D^2}{(1+\alpha)^2x^3}}
{\sqrt{x^{2-3\gamma} - k(1+\alpha)}}. \label{dydx8}
\end{equation}

	We shall assume that 
\begin{equation}
\alpha \ll 1. \label{small-alpha}
\end{equation}
We can test that approximation from the solution later and
iterate if necessary.  In that case, (\ref{x7}) and (\ref{dydx7})
become
\begin{equation}
x^{\prime} = \sqrt{x^{2-3\gamma} - k}. \label{x9}
\end{equation}
and
\begin{equation}
\frac{dy}{dx} = \frac{3kx - \frac{k}{(2a_0)^6}\frac{2D^2}{x^3}}
{\sqrt{x^{2-3\gamma} - k}}. \label{dydx9}
\end{equation}

	We notice that (\ref{dydx9}) could be integrated numerically
to obtain an approximate solution for $y(x)$.  For a closed
universe (which is the case we are considering), we can  get an
estimate of that solution, at least for small $x$.  The second
term under the radical in (\ref{x9}) and (\ref{dydx9}) is smaller
than the first term under the radical except at the point of
maximum expansion, when they are equal (when $x$ is one).  Thus,
before the universe gets too close to the point of maximum
expansion, we can neglect the second term under the radical in
(\ref{x9}) and (\ref{dydx9}) to give
\begin{equation}
x^{\prime} = x^{1-\frac{3}{2}\gamma}. \label{x10}
\end{equation}
and
\begin{equation}
\frac{dy}{dx} = 3kx^{\frac{3}{2}\gamma} -
2kE^2x^{\frac{3}{2}\gamma-4} , \label{dydx10}
\end{equation}
where
\begin{equation}
E \equiv D/(2a_0)^3 = \frac{b/r_{m1}^2}{(2a_0)^3} .
\label{Edef}
\end{equation}
We can integrate (\ref{dydx10}) to give
\begin{equation}
y = y_0 + \frac{6k}{3\gamma+2}(x^{\frac{3}{2}\gamma+1} -
x_0^{\frac{3}{2}\gamma+1}) - \frac{4k}{3}\frac{E^2}{\gamma-2}
(x^{\frac{3}{2}\gamma-3} - x_0^{\frac{3}{2}\gamma-3}) ,
\label{y10}
\end{equation}
where, without loss of generality, the constant of integration
has been chosen such that $y=y_0$ when $x=x_0$.  

	We notice that on substituting (\ref{y10}) into (\ref{x7})
and (\ref{dydx7}) the terms it gives that do not involve $b$ have
a higher power of $x$ than the dominant term.  Therefore, for
small $b$ and small $x$, it was justified to neglect the fourth
term under the radical in (\ref{x7}) and (\ref{dydx7}), since the
terms that involve $b$ are proportional to a positive power of
$b$.  

	Using (\ref{y10}) in (\ref{alpha6}) gives
\begin{equation}
\frac{\alpha^{\prime}}{1+\alpha} = 2y_0x^{-3} +
\frac{12k}{3\gamma+2}(x^{\frac{3}{2}\gamma-2} -
x_0^{\frac{3}{2}\gamma+1}x^{-3}) -
\frac{8k}{3}\frac{E^2}{\gamma-2} (x^{\frac{3}{2}\gamma-6} -
x_0^{\frac{3}{2}\gamma-3}x^{-3}) . \label{alpha10}
\end{equation}
Combining (\ref{alpha10}) with (\ref{x10}) gives
\begin{equation}
\frac{d \ln(1+\alpha)}{dx} = 2y_0x^{\frac{3}{2}\gamma-4} +
\frac{12k}{3\gamma+2} (x^{3\gamma-3}-
x_0^{\frac{3}{2}\gamma+1}x^{\frac{3}{2}\gamma-4}) -
\frac{8k}{3}\frac{E^2}{\gamma-2} (x^{3\gamma-7} -
x_0^{\frac{3}{2}\gamma-3}x^{\frac{3}{2}\gamma-4}) .
\label{dalphadx10}
\end{equation}
We can integrate (\ref{dalphadx10}) to give
\begin{eqnarray}
\ln\frac{1+\alpha}{1+\alpha_0} & = & 
\frac{4y_0}{3\gamma-6}(x^{\frac{3}{2}\gamma-3} -
x_0^{\frac{3}{2}\gamma-3}) + 
\nonumber \\
 & & 
\frac{12k}{3\gamma+2} 
\frac{(3\gamma-6)x^{3\gamma-2} -
(6\gamma-4)x_0^{\frac{3}{2}\gamma+1}x^{\frac{3}{2}\gamma-3} +
(3\gamma+2)x_0^{3\gamma-2}}
{(3\gamma-2)(3\gamma-6)} 
\nonumber \\
 & & 
- \frac{8k}{3}\frac{E^2}{\gamma-2} 
\frac{x^{3\gamma-6}-2x_0^{\frac{3}{2}\gamma-3}x^{\frac{3}{2}\gamma-3}
+ x_0^{3\gamma-6}} {3\gamma-6} , \label{alpha11}
\end{eqnarray}
where, without loss of generality, the constant of integration
has been chosen such that $\alpha=\alpha_0$ when $x=x_0$.  

	Writing (\ref{y10}) and (\ref{alpha11}) in original
variables using (\ref{xdef}), (\ref{ydef}), (\ref{Edef}), and
(\ref{alpha4}) gives 
\begin{eqnarray}
V & \equiv & \frac{r^3}{2}\frac{\dot{\alpha}}{1+\alpha} =
\frac{r_0^3}{2}\frac{\dot{\alpha}_0}{1+\alpha_0} + \nonumber \\ &
& [6k\frac{r^{\frac{3}{2}\gamma+1} -
r_0^{\frac{3}{2}\gamma+1}}{3\gamma+2} -
\frac{4k}{3}\frac{b^2}{(2a_0)^6} \frac{r^{\frac{3}{2}\gamma-3} -
r_0^{\frac{3}{2}\gamma-3}}{\gamma-2}]r_m^{1-\frac{3}{2}\gamma} ,
\label{y10b}
\end{eqnarray}
\begin{eqnarray}
1+\alpha & \approx & (1+\alpha_0) exp\left\{
\frac{2r_0^3}{3\gamma-6}\frac{\dot{\alpha}_0}{1+\alpha_0}
(r^{\frac{3}{2}\gamma-3} - r_0^{\frac{3}{2}\gamma-3})
r_m^{1-\frac{3}{2}\gamma} + \right.
\nonumber \\
 & & 
12k 
\frac{(3\gamma-6)r^{3\gamma-2} -
(6\gamma-4)r_0^{\frac{3}{2}\gamma+1}r^{\frac{3}{2}\gamma-3} +
(3\gamma+2)r_0^{3\gamma-2}}
{(3\gamma+2)(3\gamma-2)(3\gamma-6) r_m^{3\gamma-2} } 
\nonumber \\
 & & 
\left. - \frac{8k}{3}\frac{b^2}{(2a_0)^6} 
\frac{r^{3\gamma-6}-2r_0^{\frac{3}{2}\gamma-3}r^{\frac{3}{2}\gamma-3}
+ r_0^{3\gamma-6}} {(\gamma-2)(3\gamma-6)r_m^{3\gamma-2} }
\right\}, \label{alpha11b}
\end{eqnarray}


\end{document}